\documentclass[a4paper]{llncs}

\usepackage{amssymb}
\usepackage{graphicx}

\usepackage{url}
\usepackage{amsmath}

\usepackage{subfig}
\usepackage[labelfont=bf]{caption}

\usepackage{caption}
\DeclareCaptionLabelSeparator{dot}{. }
\begin{document}
	
	\mainmatter  % start of an individual contribution

	\title{The Parallel Monte Carlo Algorithm Implementation on GPU for the Systems with an Ising Hamiltonian under the Condition of a Constant Charge Density}
	\titlerunning{Metropolis algorithm for Ising model with constant charge density}
	\author{K.S. Budrin
		\and V.A. Ulitko
		\and A.A. Chikov
		\and Yu.D. Panov
		\and A.S. Moskvin}
	
	\institute{Ural Federal University, Ekaterinburg, Russia}
	
	% * <vasily.ulitko@gmail.com> 2018-02-19T14:22:57.288Z:
	\maketitle
	
	\begin{abstract}
		This paper is devoted to computational algorithms designed to describe the classical Ising magnet in some specific cases when an additional macroscopic restriction in form of constant charge density exists in the system. We developed and implemented a parallel algorithm for modeling such a systems on GPU with CUDA technology. This work focuses on technical aspects of implementing the algorithm.\\
		
	\end{abstract}
	% \keywords{Monte-Carlo, Metropolis, cuprate superconductor, Ising, GPU, CUDA}

	\section{Introduction}
	
	For developing computer models of thermodynamic systems, the Metropolis algorithm described by N. Metropolis and S.Ulam is often used \cite{metropolis-ulam,monte-carlo-guide}. Recently, the implementation of this algorithm for parallel computations on the GPU is of interest. We consider an Ising magnet with the condition of constant charge density. Such a situation exists when describing the competition between charge and magnetic ordering in superconducting cuprates \cite{ground-state}.
	The article describes a parallel Metropolis algorithm that implements this constraint.
	The article has the following structure.
	In the theoretical part of the work, we describe the mathematical model of the described system and formulate the problem.
	In the third part, we present the algorithm developed by us, and also describe the method of parallelizing it on a GPU using the CUDA technology.
	In Sections 4 and 5, we present the simulation results and analyze the performance of the algorithms used.
	
	\section{Theory}
	When studying high-temperature superconductivity in cuprates, there is the problem of modeling systems with a competition of magnetic and charge ordering. We considered an Ising magnet with a fixed total charge (fixed charge density). Physical aspects are not of great importance for this article. This article is devoted to algorithms and problems of their implementation, so we will go directly to the description of the mathematical model. Detailed description of the physical model can be found in \cite{cuprate}. 
	
	\subsection{Mathematical Model}
	We consider a square two-dimensional lattice of size $L\times L$. We let $S$ as charge on the site, and let $s$ as spin projection to $z$ axis. Each site can be in one of 4 possible states: two magnetic states (spin projection $s=\pm \frac{1}{2}$) and two charge ones ($S =\pm 1$). For magnetic states $S=0$.
	Energy of this system:
	
	\begin{equation}\label{eq-energy}
		E= \Delta \sum_i^{\phantom{N}} S_{i}^2 
		+ V \sum_{\left\langle ij\right\rangle} S_{i} S_{j} 
		+ J \sum_{\left\langle ij\right\rangle} (1-S_i^2)s_i s_j (1-S_j^2),
	\end{equation}
	where $\Delta$ and $V$ are model parameters which are related to charge coupling, and $J$ is the spin-spin coupling constant.
	The summation is over all $N$ lattice sites. $\sum\limits_{\left\langle ij\right\rangle} $ is for summation over the nearest neighbors (4 for each site).
	The system has periodic boundary conditions.
	In addition, the system is constrained:
	\begin{equation}\label{eq-charge-constrain}
		\sum_{i} S_i = n \, N = const,
	\end{equation}
	to ensure the constancy of the charge density.
	
	If we let $\sigma_i = 2(1-S_i^2)s_i $, then we can re-write equation (\ref{eq-energy}) in equivalent form:
	\begin{equation}\label{eq-energy-sigma}
		E= \Delta \sum_i^{\phantom{N}} S_{i}^2 
		+ V \sum_{\left\langle ij\right\rangle} S_{i} S_{j} 
		+ \frac{J}{4} \sum_{\left\langle ij\right\rangle} \sigma_i\sigma_j,
	\end{equation}
	where $\sigma_i$ describes a magnetic state and can take on values $\sigma=\pm 1$. Equation (\ref{eq-energy-sigma}) is more convenient for calculations and we will use this form further.
	
	\subsection{Research Objective}
	The constrain (\ref{eq-charge-constrain}) reduces the system's degrees of freedom by 1. So the classical Metropolis algorithm must be modified. One way to fix the total charge is to include an additional term in the energy expression:  $-\mu\sum\limits_i S_i  $, where $\mu$ is the chemical potential of the system. In the course of the algorithm, the value $\sum\limits_i s_i $ is controlled by automatically adjusting the parameter $\mu$. In fact, here we use the penalty function method used for solving constrained optimization problems\footnote{When using the classical method of penalty functions, an additional term should be taken in the form $\mu(\sum\limits_i s_i-nN)  $. In this case, the penalty coefficient $\mu$ will not have the meaning of the chemical potential of the system.}. A similar approach was used, for example, in the works \cite{filipp,hard-core}. The implementation of this method was carried out by our colleagues from the Institute of Physics of Metals. Their program allowed us to observe the evolution of the system, and the results were qualitatively consistent with theoretical concepts.
	
	However, when calculating the thermodynamic characteristics of the system (heat capacity and magnetic susceptibility), some problems were identified.
	In particular, the heat capacity of the system tends to infinity at $T \to 0$.
	We found it the following explanation.
	According to the statistical definition of the heat capacity
	\cite{newman}, 
	\begin{equation*}
		C(T) = \frac{1}{N} \frac{\langle  E^2 \rangle -\langle  E \rangle^2}{k T^2},
	\end{equation*}
	where $E$ is the energy of the system, $N$ is the number of the sites, $k$ is the Boltzmann constant, and $\langle ...\rangle $ is the statistical mean. Quantity $\langle  E^2 \rangle -\langle  E \rangle^2$ is the energy dispersion of the system.
	A feature in the heat capacity can be a consequence of the penalty function -- the system experiences "fluctuations" that do not disappear at zero temperatures. As a result, there is always a nonzero energy variance and, consequently, $C(T) \to \infty$ when $T \to 0$.
	Thus, this method doesn't suit for calculating the thermodynamic characteristics.
	Our purpose was to create such a modification of the Metropolis algorithm, which would guarantee the fulfillment of the condition (\ref{eq-charge-constrain}) at each step. At the same time, the possibility of parallelizing the algorithm should be conserved.
	
	\section{Implementation}
	
	\subsection{The Metropolis Algorithm with "Coupled Pairs"}
	
	The basis is the assumption that from any state of our configuration space (bounded by the condition (\ref{eq-charge-constrain})) we can reach any other state of this space using a sequence of pairwise changes in the state of sites.
	Then we can reduce the problem of fixing the total charge to the problem of conservation of charge on a pair. 
	This is achieved in two stages:
	\begin{enumerate}
		\item Obtaining a configuration with a given charge from an arbitrary state. This can be done by turning the spins in the desired direction at random lattice sites.
		\item then using  the Metropolis algorithm with "coupled pairs".
	\end{enumerate}
	
	\textbf{Algorithm}
	
	At each step of the algorithm, a pair of sites $a$ and $b$ -- "coupled pair" -- are randomly selected. Then we calculate the total charge $ q = s_a + s_b $ for this pair. In the present model, $s_a$ and $s_b$ can take one of 3 possible values: $\pm 1$ or 0. In this case, if $S_i$ = 0, then $s_i$ can be equal to $\pm\frac{1}{2}$ (magnetic state).
	All possible configurations of pairs for each value of $q$ are:
	\par $q = \pm 2: (\pm 1,\pm 1) $ -- 1 configuration;
	\par $q = \pm 1: \left( \pm1,\frac{1}{2}\right),\left(  \pm1,-\frac{1}{2} \right),\left( \frac{1}{2}, \pm1\right),\left( -\frac{1}{2} , \pm1\right)$ -- 4 configurations;
	\par $q = 0: \left(  \pm1, \mp 1 \right),\left(\pm \frac{1}{2}, \pm\frac{1}{2},\right),\left( \pm\frac{1}{2} ,\mp \frac{1}{2},\right)$ -- 6 configurations.\\
	% * <vasily.ulitko@gmail.com> 2018-02-19T09:28:09.193Z:
	% 
	% путаница с числом конфигураций?
	% 
	% ^.
	If $ q = \pm 2 $, the configuration can not be changed, and we can go directly to the next step. In other cases, in order to preserve the total charge, it is sufficient to randomly select from the corresponding set one two-site configuration and calculate the energy change $dE$ for this pair. 
	First of all, the algorithm calculate all possible configuration of pair fir each value of the charge for further use.
	First of all, a set of possible configurations of pairs of sites (for each possible value of the charge) is stored in memory for further use. Otherwise, this algorithm does not differ from the classical Metropolis algorithm. When implementing a parallel algorithm on CUDA, it is advisable to store a set of two-site configurations in the constant memory of the video card. This ensures the maximum possible efficiency of interaction with memory.
	
	When calculating $dE$, one should distinguish between the case of distant and neighboring sites. Let us write the difference between the energies of the states 1 and 0 for the sites $a$ and $b$:
	$$dE = E_1 - E_0 = \Delta E_{\Delta} + V E_V + J E_J.$$
	Then in the case of distant sites from (\ref{eq-energy-sigma}) we get:
	\begin{equation}
		\begin{split}
			& E_{\Delta} = S_{a1}^2 + S_{b1}^2 - S_{a0}^2 -S_{b0}^2 ,\\
			& E_V = (S_{a1} - S_{a0}) \bigg(\sum\limits_{\langle a \rangle} S_i  - \sum\limits_{\langle b \rangle} S_i \bigg) ,\\
			& E_J = (\sigma_{a1}-\sigma_{a0}) \bigg(\sum\limits_{\langle a \rangle}\sigma_i - \sum\limits_{\langle b \rangle}\sigma_i\bigg).\\
		\end{split}
	\end{equation}
	
	In the case of neighboring sites, the expressions for $ E_V $ and $ E_J $ change to:
	\begin{equation}
		\begin{split}
			& E_V = (S_{a1} - S_{a0}) \bigg(\sum\limits_{\langle a \rangle} S_i  - \sum\limits_{\langle b \rangle} S_i \bigg) - (S_{a1} - S_{a0})^2 ,\\
			& E_J = (\sigma_{a1}-\sigma_{a0}) \bigg(\sum\limits_{\langle a \rangle}\sigma_i - \sum\limits_{\langle b \rangle}\sigma_i\bigg)  + (\sigma_{a1} - \sigma_{a0})(\sigma_{b1} - \sigma_{b0}).\\
		\end{split}
	\end{equation}
	
	The symbol $\sum \limits _ {\langle ... \rangle}$ denotes summation over the nearest neighbors of the corresponding site. In the case of neighboring sites $ a $ and $ b $, the sum $ \sum \limits_{\langle a \rangle} $ includes $ S_ {b0} $, and the sum $ \sum\limits_{\langle b \rangle} $ includes $ S_{a0}$. In this expression the condition for constant charge on the pair is taken into account: $ S_{a1} + S_ {b1} = S_{a0} + S_{b0} $.

	\subsection{Parallel Algorithm}
	\begin{figure}[h]
		\center{\includegraphics[width=0.5\linewidth]{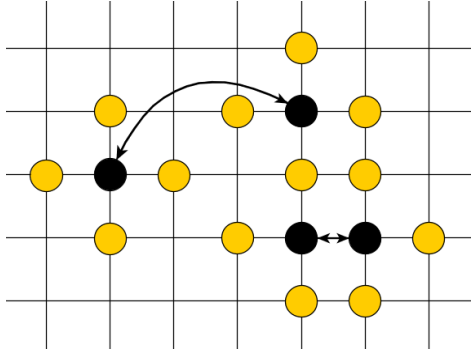}}
		\caption{
			Covering of a map for two pairs of sites
		}
		\label{dia-pare}
	\end{figure}
	
	Considering only the nearest neighbors, the parallel version of the Metropolis algorithm can be organized as follows. The lattice is divided into two sublattices corresponding to an even and odd sum of the coordinates of the sites (similar to the black and white cells of the chessboard). The sublattices are processed in turn. The computation of  $dE$ for each sublattice site does not depend on the states of other sites of this sublattice. Therefore, the sublattice can be processed in parallel, giving a separate stream to each site. A similar parallel algorithm was used in \cite{filipp}. However, such variant is not correct  for an algorithm with coupled pairs. Therefore, we found another way of parallelizing, which consists in drawing up "maps" of future calculations. Map is referred to a randomly generated list of site pairs that do not interact with each other. The size of the map (the number of such pairs) is fixed. A "covering" of a map is a collection of sites from a given map, as well as their neighbors (Fig. \ref{dia-pare}). Thus, the map contains a list of coupled pairs that can be processed in parallel. 
	
	The map size is chosen according to the following considerations:
	\begin{enumerate}
		\item Such a number of pairs must exist under any generation conditions.
		\item Coverage should have the maximum possible area.
	\end{enumerate}
	Proceeding from the general geometry of the problem, the number of pairs is chosen equal to $ N / 10 $, where $ N $ is the total number of lattice sites. The algorithm for constructing such a map is described below.

	\textbf{Algorithm for building a map}
	\begin{enumerate}
		\item Declare a two-dimensional $ L\times L$ array of elements of type bool, fill it $false$;
		\item Randomly select one item. If the value of the element is true, repeat until there is an element witch value is $false$.
		\item Change the value of the element from point 2 to true, saving the pair of its coordinates $ (x_a, y_a) $;
		\item Repeat steps 2-3 and obtain the second pair of coordinates $ (x_b, y_b) $;
		\item Add the obtained pair of sites to the map;
		\item Set the values of nearest neighbors obtained pair to $true$.
		\item Repeat steps 2-6 until the map is full.
	\end{enumerate}
	
	\begin{figure}[h]
		\center{\includegraphics[width=1\linewidth]{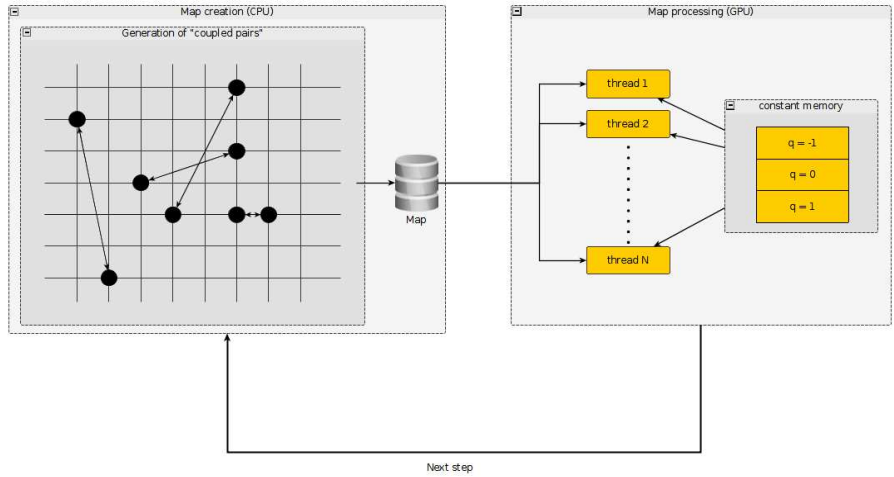}}
		\caption{Creation and processing of "map"}
		\label{map-gen-proc}
	\end{figure}
	
	During the testing of the algorithm, we came to the conclusion that it is expedient to generate the maps on the CPU. The effective bandwidth of the CPU interface $ -> $ GPU is higher the more memory is transferred at a time. Because of this, we realized the following scheme. A large number of maps are generated on the CPU. The array of maps transmitted to the video card is processed sequentially. The optimal size of the transmitted array of maps depends on the hardware configuration, and it must be selected individually. Each map is processed by a video card in parallel, with each of the threads processing its pair of sites according to the Metropolis algorithm with coupled pairs. Schematically this process is depicted in Figure \ref{map-gen-proc}.
	
	\begin{figure}[h]
		\center{\includegraphics[width=0.5\linewidth]{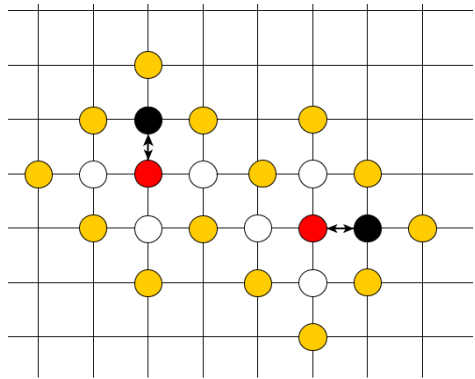}}
		\caption{
			Covering of a map for two pairs of close sites
		}
		\label{dia-pare-close}
	\end{figure}
	
	For multi-core processors, the generation of an array of maps can also be performed in parallel. For this purpose, we used the OpenMP standard.
	
	When calculating the thermodynamic characteristics of the system, the algorithm can be run on several copies of the lattice at the same time, then averaging over these copies. This ensures a better convergence of the results. In our implementation, the generated array of maps was sent to all copies of the lattice, and all copies were processed in parallel (as far as possible to fit the capabilities of the video card).
	
\begin{figure}[h]
		\begin{minipage}[h]{0.49\linewidth}
			\center{\includegraphics[width=0.9\linewidth]{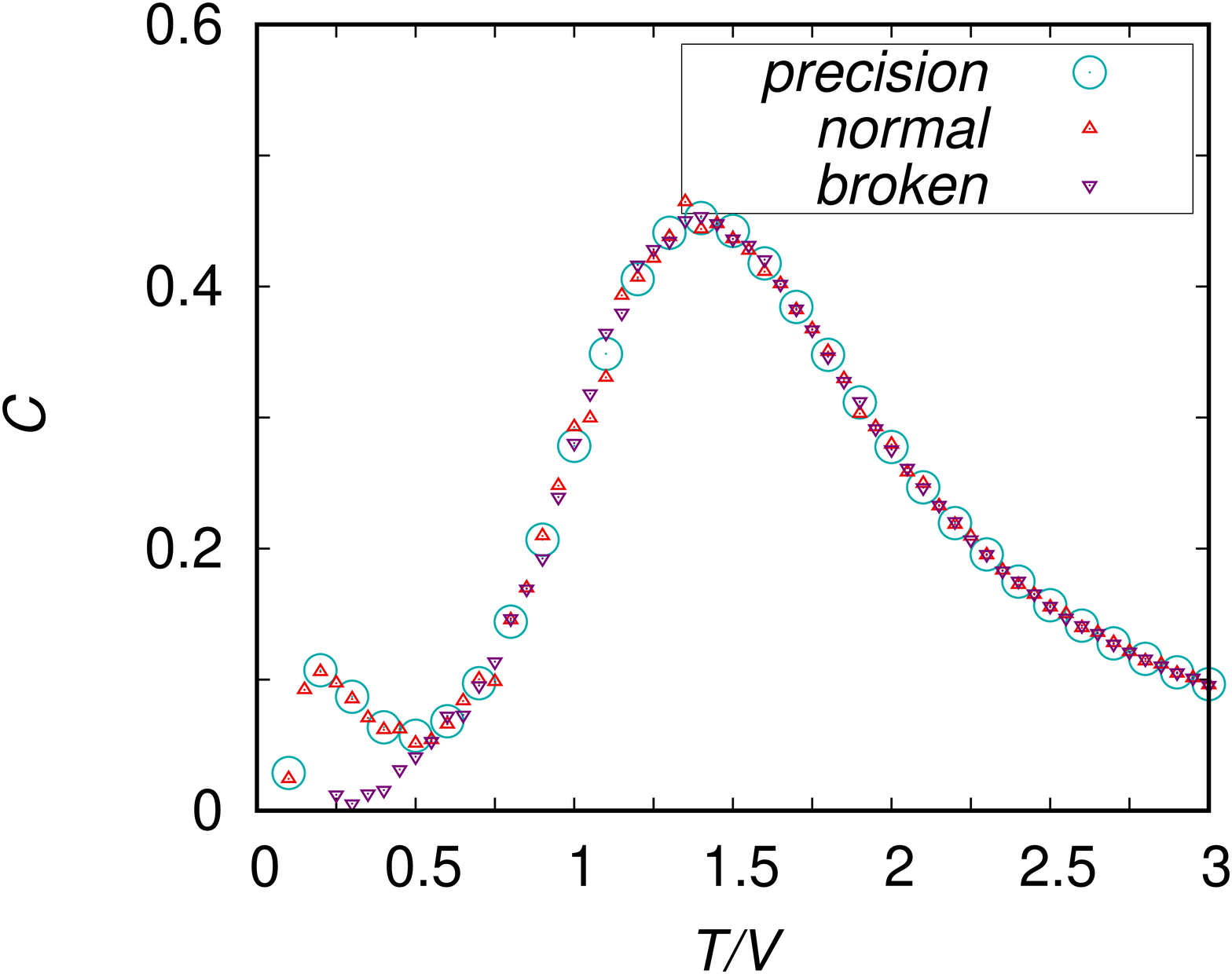} \\Lattice $4\times4$}
		\end{minipage}
		\hfill
		\begin{minipage}[h]{0.49\linewidth}
			\center{\includegraphics[width=0.9\linewidth]{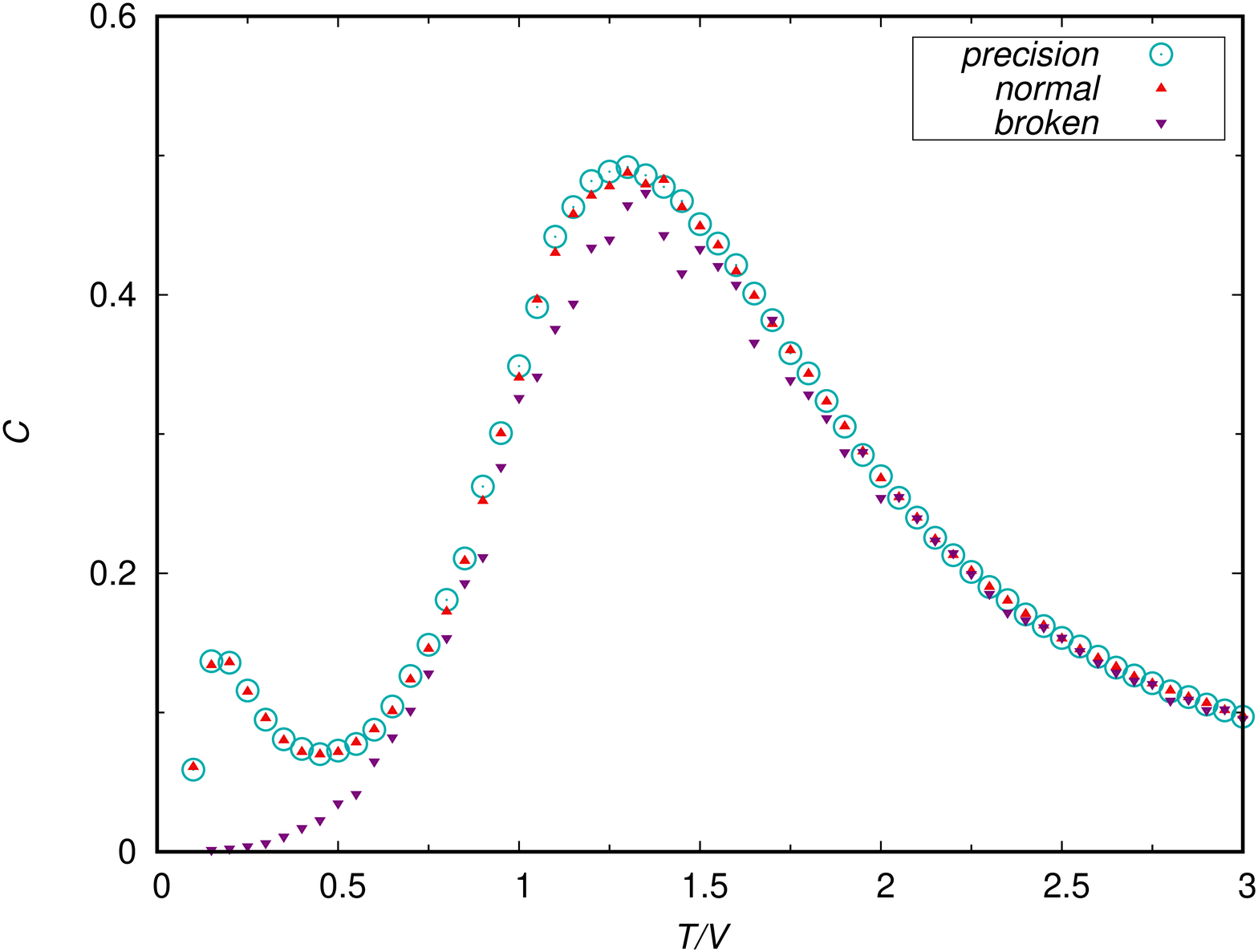} \\ Lattice $16\times16$}
		\end{minipage}
		\caption{Comparison of the results obtained by different algorithms on the example of the temperature dependence of the heat capacity. The "reference" solution on Scala is represented by round markers. Triangular markers represent the result of the operation of the algorithm with the choice of random pairs and an erroneous algorithm with "short range" respectively}
		\label{fig-tests}
	\end{figure}

	It is especially important to note the possibility of breaking the ergodic algorithm by the simulated system.
	In the first version of our algorithm, the coupled pair was always located on adjacent sites: one site was randomly selected, the other chosen among it's neighbors (Fig. \ref{dia-pare-close}). During the test on lattice $4\times4$ results of the test didn't match with the precision results in the low-temperature region. In particular, there was no second peak in the temperature dependence of the heat capacity (Fig. \ref{fig-tests}). An analysis of these results showed that the reason for their formation is the "short-range" property of the algorithm - some of the system states are unattainable during the operation of the algorithm. Thus, the idea of the nearest neighbors should be abandoned in favor of a completely random choice of coupled pairs.
	
	\begin{figure}[h]
		{\includegraphics[width=1.0\linewidth]{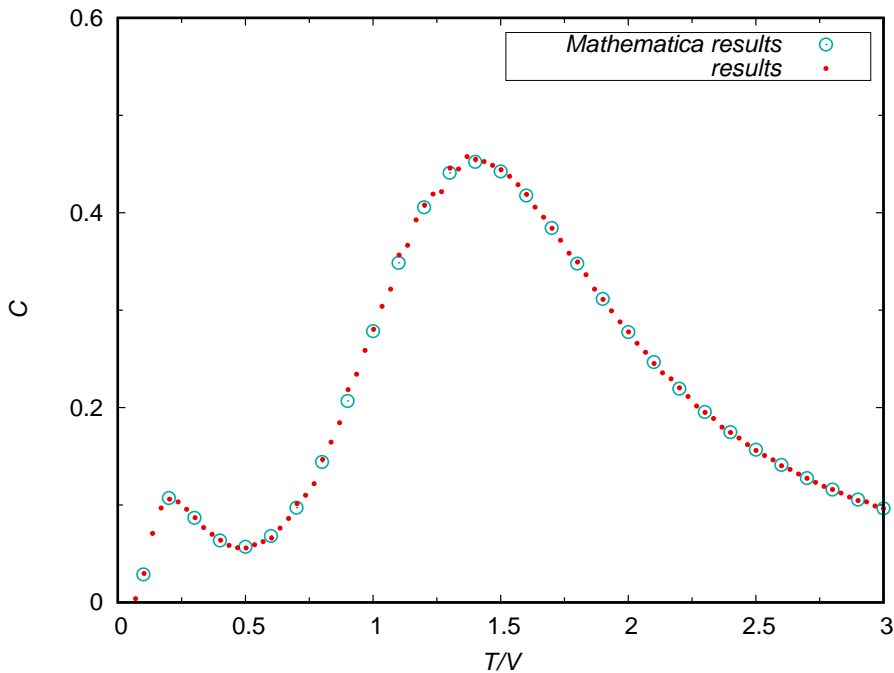}}
		\caption{
			Comparison of the exact solution with the results obtained by a single-threaded program on Scala for a $4 \times 4$ lattice
		}
		\label{scala-verify}
	\end{figure}

	\subsection{Testing}
	During the development of the algorithm, it was tested at each step. In the paper \cite{ground-state} the analytical results for our model, obtained in the mean-field approximation, are described. However, they are of a qualitative nature and are not suitable for an accurate verification of the numerical solutions obtained. For these purposes, we used the exact solution of our problem for the $ 4 \times4 $ lattice calculated in the software package Wolfram Mathematica. All implementations of our algorithm were compared with this reference solution. To test the operation of the algorithm on large lattice, a linear (single-threaded) program in Scala was written. The choice of Scala was due to the support of long arithmetic, which avoid an error associated with the rounding of floating-point numbers. The Scala program showed a high degree of compliance with the exact solution and was used to test the parallel program on CUDA (Fig. \ref{scala-verify}). In Fig. \ref{fig-tests} we present the results of comparing different versions of the algorithm.
	
	\begin{figure}[h]
		\centering
		\includegraphics[width=0.9\textwidth]{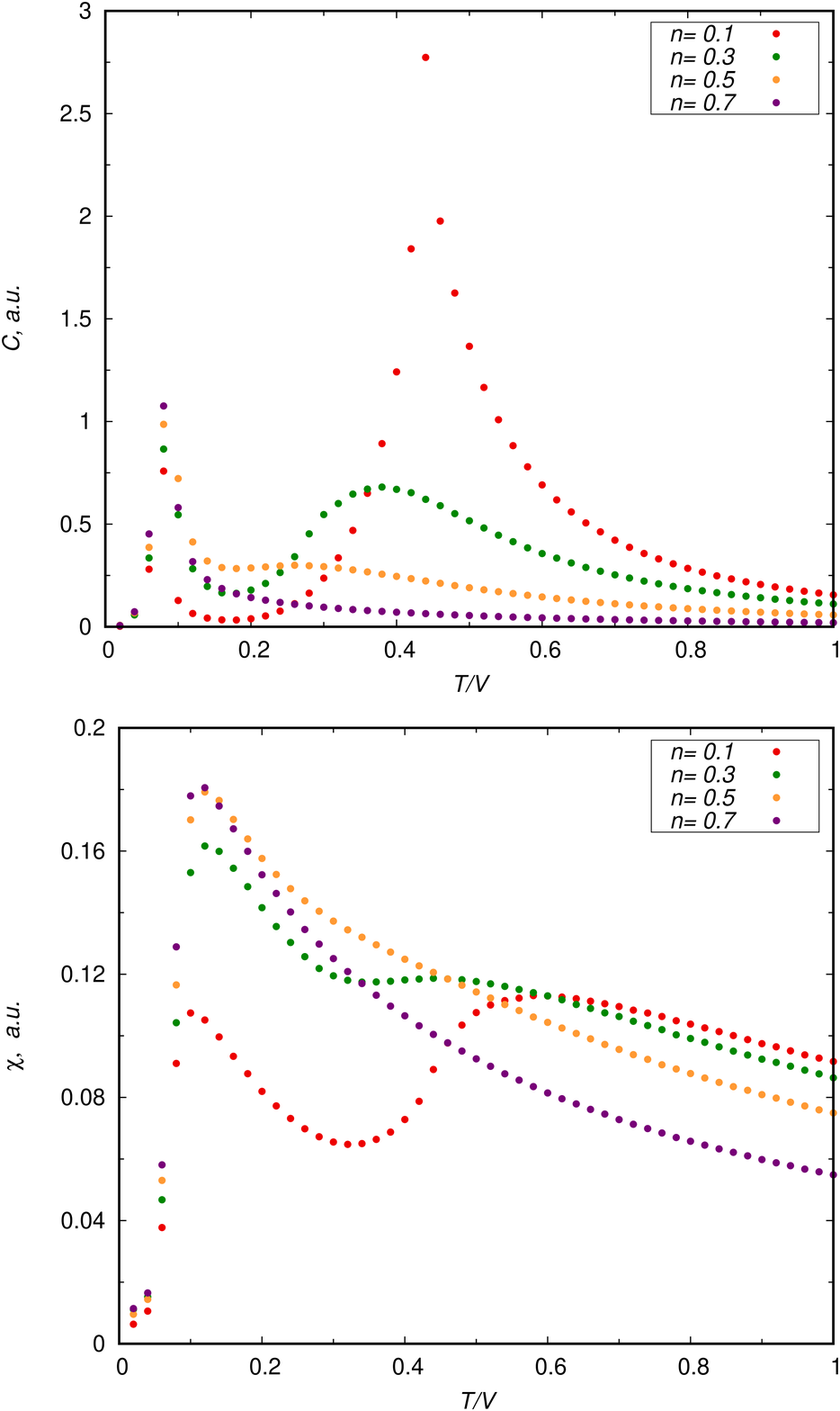}
		\caption{Temperature dependencies of specific heat and magnetic susceptibility for various $n$}
		\label{mism-plot-2}
	\end{figure}

	\begin{figure}[h!]
		\center{\includegraphics[width=1\linewidth]{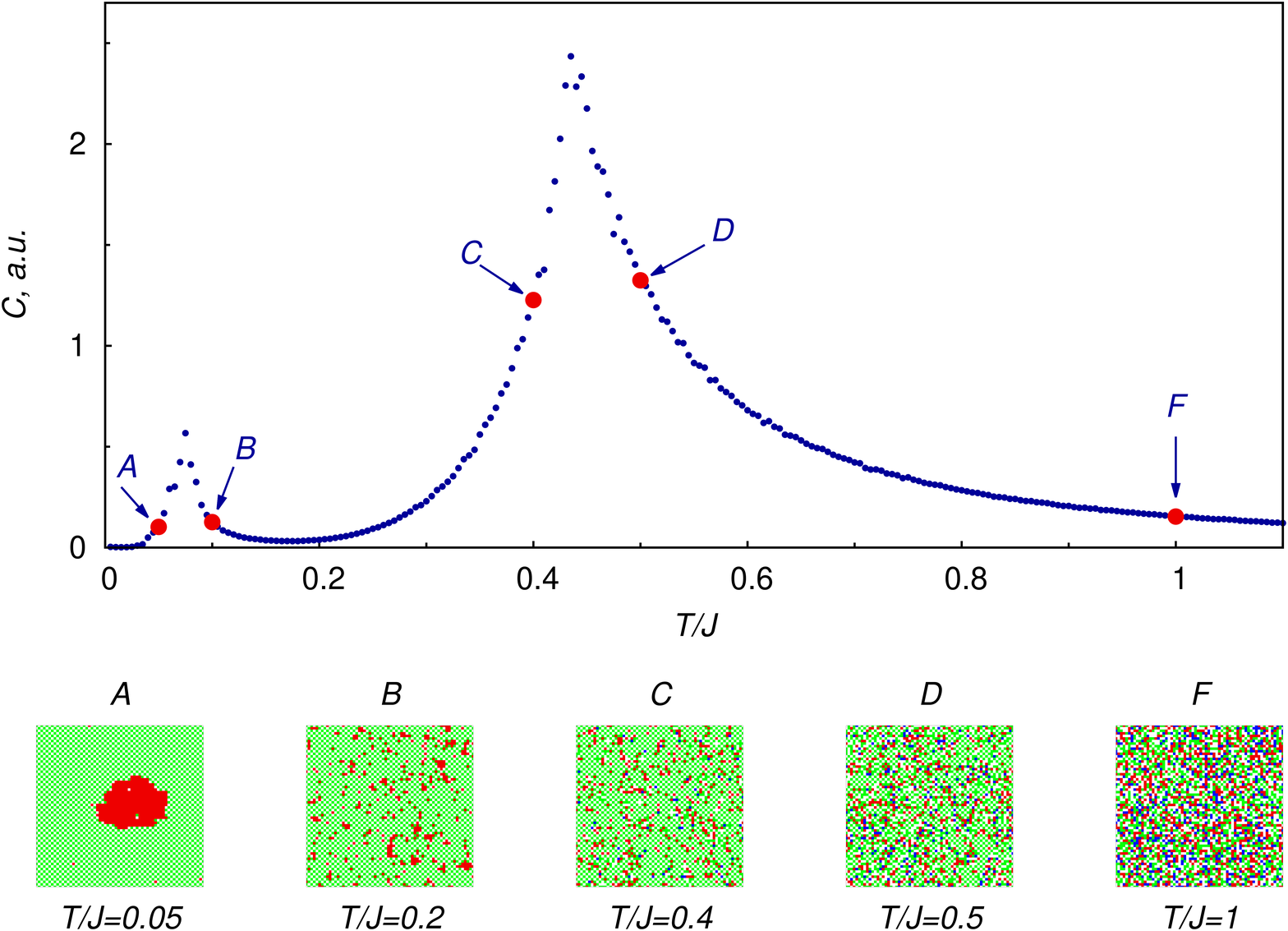}}
		\caption{Formation of a charge drop. Red and blue colors correspond to charge states with $ S = 1 $ and $ S = -1 $, respectively, green and white colors correspond to spin states with $ s = \frac{1}{2} $ and $s = -\frac{1}{2} $, respectively}
		\label{drop-creation}
	\end{figure}
	
	\section{Results}
	As the result of the work, the thermodynamic characteristics of the system were calculated, as well as the form of the states of the system in the vicinity of the phase transitions. The figure \ref{mism-plot-2} shows the temperature dependence of the thermodynamic parameters for different concentrations of the doped charge. An original result was obtained, connected with the competition of charge and magnetic orders.
	In configurations with a strong magnetic exchange ($ J >> V $), a situation is observed when the total charge of the system accumulates in one place awhile the rest of the lattice remains antiferromagnetically ordered. Thus the "drop" is formed.
	The nature and physical significance of this state are discussed in \cite{cuprate}.
	In Fig. \ref{drop-creation} we show the process of formation of a "drop" for the lattice of size $64\times 64$.

	\section{Performance}
\begin{figure}[h]
		\begin{minipage}[h]{0.49\linewidth}
			\center{\includegraphics[width=0.9\linewidth]{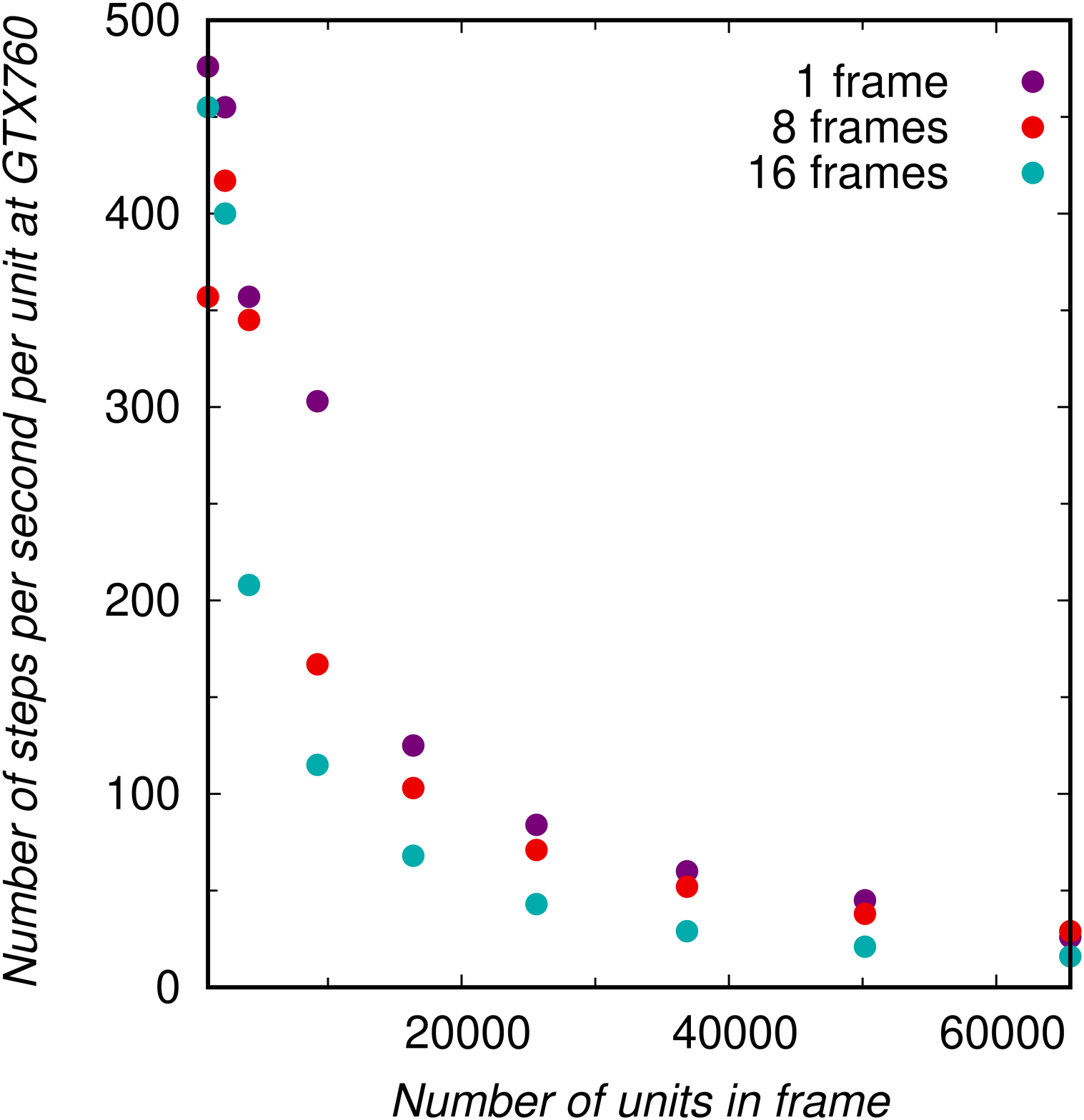} \\Tesla K10}
		\end{minipage}
		\hfill
		\begin{minipage}[h]{0.49\linewidth}
			\center{\includegraphics[width=0.9\linewidth]{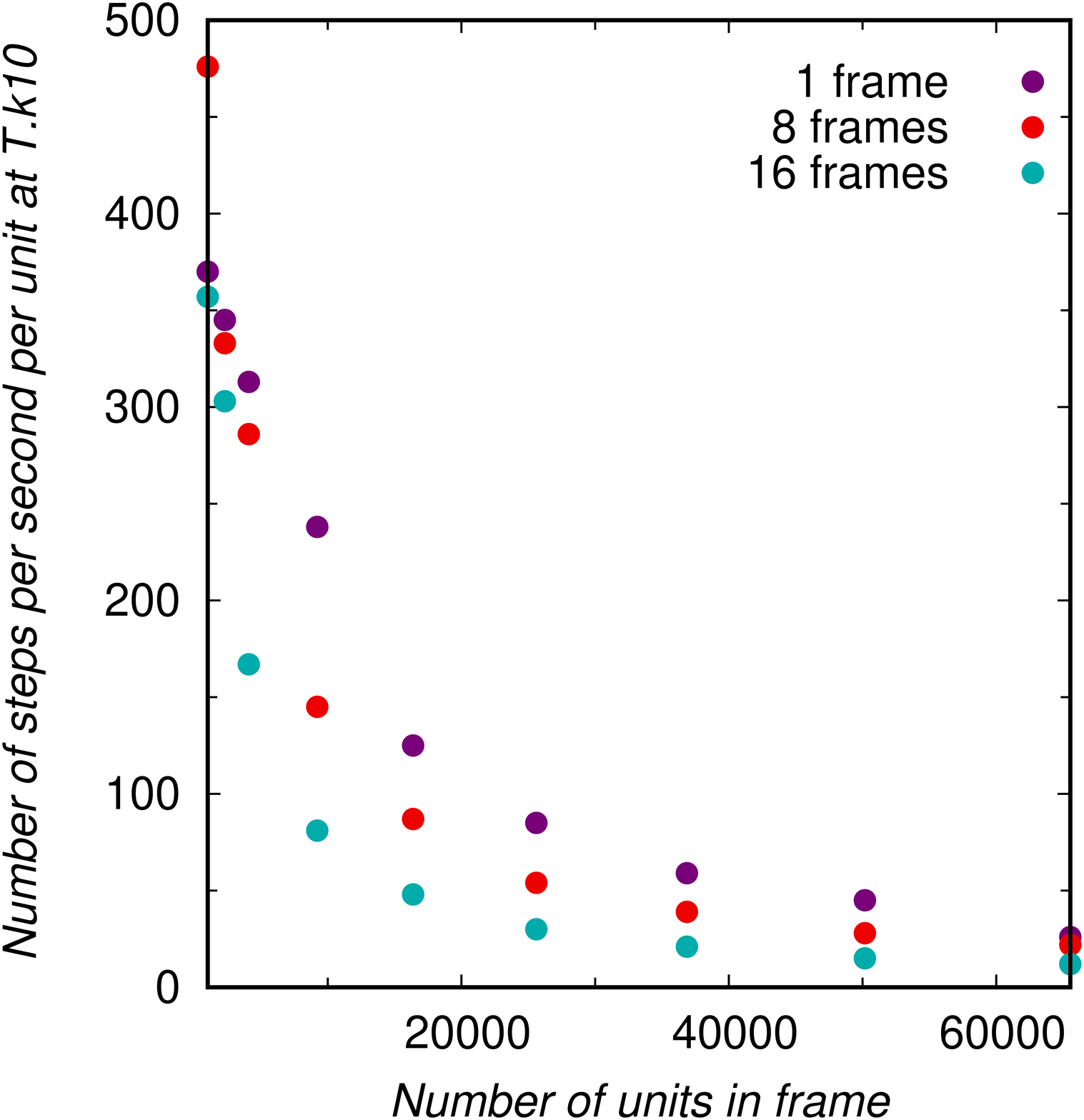} \\ GTX 760}
		\end{minipage}
		\caption{The dependence of the speed of the algorithm on the number of lattice sites}
		\label{performance}
	\end{figure}

	For performance analysis, we carried out the test calculations with a fixed number of steps. The step of the algorithm is the processing of all the elements of one map. The performance parameter was the average number of steps per second. The calculations were made for several systems with different hardware characteristics. In Fig. \ref{performance} there are dependencies of the average number of steps per second on the number of sites for a various number of copies of the lattice.
	
	\section{Conclusion}
	In this paper, we described a mathematical model of an Ising magnet with a conserved charge density. A modification of the Metropolis algorithm for such a system was presented. The obtained results coincided with the exact solution of the problem for the $ 4 \times4 $ lattice. We also proposed a technique for parallelizing this algorithm on GPU using CUDA technology. A feature of the parallel version of the algorithm is the dependence of the performance on the characteristics of the CPU and GPU.
	
	The performance analysis showed that the running time of the algorithm is not proportional to the number of lattice sites. This time is less than it would be in the case of proportional dependence up to some critical number of sites. We tested the algorithm on video cards Nvidia Tesla K10 and GeForce GTX 760 and this trend was maintained up to the $ 256 \times256 $ lattice size. Thus, the algorithm remains effective for $ 256 \times 256 $ and larger sizes of lattice, as well as simultaneous calculation of several copies of the lattice.

\end{document}